\documentclass[floats,floatfix,showpacs,amssymb,prd,twocolumn,superscriptaddress,nofootinbib,nolongbibliography,reprint]{revtex4-2}

\usepackage{amssymb,amsmath,verbatim,mathtools,needspace,enumitem,etoolbox,graphicx,physics,microtype,afterpage,bigints,gensymb,tabularx,xspace}
\usepackage{siunitx}
\usepackage[dvipsnames, usenames]{xcolor}
\definecolor{linkcolor}{rgb}{0.0,0.3,0.5}
\definecolor{dodgerblue}{HTML}{1E90FF}
\usepackage[unicode, colorlinks=true, linkcolor=linkcolor, citecolor=linkcolor, filecolor=linkcolor,urlcolor=linkcolor, pdfusetitle]{hyperref}
\usepackage[all]{hypcap}
\usepackage[T1]{fontenc}
\usepackage[utf8]{inputenc}
\usepackage{orcidlink}
\usepackage{ulem}

\renewcommand{\vec}[1]{\mathbf{#1}}

\makeatletter
\newcommand*{\balancecolsandclearpage}{\close@column@grid \cleardoublepage \twocolumngrid}
\makeatother

\newcommand\masout{\bgroup\markoverwith{\textcolor{purple}{\rule[0.5ex]{3pt}{0.9pt}}}\ULon}

\begin{document}

\title{
Impact of Spacecraft Orbit Uncertainties and Velocity Mismodeling on the LISA Gravitational-Wave Response
}

\author{Lorenzo Speri
\orcidlink{0000-0002-5442-7267}}\email{lorenzo.speri@esa.int}
\affiliation{European Space Agency (ESA), European Space Research and Technology Centre (ESTEC), Keplerlaan 1, 2201 AZ Noordwijk, the Netherlands}
\affiliation{Leiden Observatory, Leiden University, P.O. Box 9513, 2300 RA Leiden, the Netherlands}

\author{Olaf Hartwig}
\affiliation{Max-Planck-Institut f\"ur Gravitationsphysik (Albert-Einstein-Institut), Callinstraße 38, 30167 Hannover, Germany}
\affiliation{Leibniz Universit\"at Hannover, Institut f\"ur Gravitationsphysik, Callinstraße 38, 30167 Hannover, Germany}

\author{Waldemar Martens}
\affiliation{European Space Agency (ESA), European Space Operation Centre (ESOC), Robert-Bosch Straße 5, 64293 Darmstadt, Germany}

\author{Oliver Jennrich\orcidlink{0000-0002-8158-2668}}
\affiliation{European Space Agency (ESA), European Space Research and Technology Centre (ESTEC), Keplerlaan 1, 2201 AZ Noordwijk, the Netherlands}

\author{Eric Joffre}
\affiliation{European Space Agency (ESA), European Space Research and Technology Centre (ESTEC), Keplerlaan 1, 2201 AZ Noordwijk, the Netherlands}

\author{Michele Armano
\orcidlink{0009-0009-0584-9859}}
\affiliation{European Space Agency (ESA), European Space Research and Technology Centre (ESTEC), Keplerlaan 1, 2201 AZ Noordwijk, the Netherlands}

\author{Martin Hewitson}
\affiliation{European Space Agency (ESA), European Space Research and Technology Centre (ESTEC), Keplerlaan 1, 2201 AZ Noordwijk, the Netherlands}

\author{Nora Lützgendorf}
\affiliation{European Space Agency (ESA), European Space Research and Technology Centre (ESTEC), Keplerlaan 1, 2201 AZ Noordwijk, the Netherlands}

\date{\today}

\begin{abstract}

The Laser Interferometer Space Antenna (LISA) is a space-based gravitational wave observatory that consists of three spacecraft in a near-equilateral triangular formation.
The spacecraft orbits are typically assumed to be perfectly known in LISA data analysis studies, but in reality, the orbit determination process introduces uncertainties in the spacecraft positions and velocities.
In this work, we investigate how these uncertainties propagate into the LISA detector output and the impact of neglecting the spacecraft velocities.
We quantify these errors in the knowledge of the LISA response using mismatches and discuss the implications for gravitational wave data analysis.
We find that spacecraft orbit uncertainties impact the LISA response knowledge at high frequencies with worst mismatch below $10^{-7}$.
The effect of neglecting the spacecraft velocities is largest at frequencies around $10^{-4}$ Hz with mismatches of order $10^{-4}$.
For a galactic binary with frequency $10^{-4}$ Hz and SNR=200 observed for one year, we find that neglecting the spacecraft velocities in the response leads to less than 1-$\sigma$ biases in the parameter estimates.
This work provides the first characterization of how errors in the LISA gravitational wave response propagate from gravitational wave strain through detector output to estimated parameters.

\end{abstract}

\maketitle


\section{Introduction}
The Laser Interferometer Space Antenna (LISA) is a space-based gravitational wave (GW) observatory scheduled for launch in the mid-2030s by the European Space Agency (ESA), designed to detect GWs in the millihertz frequency band~\cite{2024arXiv240207571C}.
LISA consists of three spacecraft forming a near-equilateral triangle with arm lengths of approximately 2.5 million kilometers, trailing Earth in a heliocentric orbit.
The mission will open a unique observational window on a rich variety of sources, including merging massive black hole binaries (MBHBs), extreme mass-ratio inspirals (EMRIs), and compact galactic binaries, as well as stochastic GW backgrounds of cosmological origin.

The measurement principle of LISA relies on heterodyne laser interferometry between free-falling test masses aboard each spacecraft. 
Because laser frequency noise would otherwise overwhelm the GW signal by many orders of magnitude, the raw one-way phase measurements are post-processed using Time-Delay Interferometry (TDI), which synthesizes virtual equal-arm interferometers by linearly combining time-shifted single-link measurements~\cite{1999ApJ...527..814A,2021LRR....24....1T}.
The mapping from the incoming GW strain to the fractional-frequency measurement depends critically on the spacecraft positions and velocities through the arm lengths, unit vectors between spacecraft, and the Doppler phases accumulated during photon propagation.
The analytic form of this gravitational wave single-link response was established by Refs.~\cite{1998PhRvD..57.7089C,2003PhRvD..67b9905C}, and its Fourier-domain representation was developed in Ref.~\cite{2018arXiv180610734M}.
Extensions to account for the fully relativistic Doppler boosting of the response were derived in Ref.~\cite{2025arXiv250910038V}, and the finite-arm-length effects on parameter estimation for MBHBs were investigated in Ref.~\cite{2002PhRvD..66l2001S}.
The LISA orbit has been designed and optimised to minimise flexing of the constellation~\cite{2021JAnSc..68..402M}, and sensitivity curves encoding the detector response have been computed in Refs.~\cite{2000PhRvD..62f2001L,2012CQGra..29l4015V}.

Accurate and computationally efficient evaluation of the LISA response is essential for GW data analysis, since parameter estimation and source localization require millions of likelihood evaluations, each involving a full response computation.
Significant effort has therefore been devoted to accelerating response calculations, for example through GPU-accelerated implementations~\cite{2022PhRvD.106j3001K,2025PhRvD.112f4017G,2026arXiv260212011N}, hybridization and finite-difference techniques~\cite{2025PhRvD.112d4020V}, and fast methods tailored to galactic binaries~\cite{2024PhRvD.110h2001R}.
Beyond speed, the fidelity of the response model is equally important: systematic errors introduced by approximations in the orbital model can bias parameter estimates and degrade source localization.
In this context, the impact of instrumental calibration uncertainty on LISA science has been assessed in Ref.~\cite{2022PhRvD.106b2003S}, and trajectory simulation software for navigation and orbit determination has been developed in Refs.~\cite{Marchese2025,godot,midas}.

Despite these advances, a key question remains unanswered: \textit{how do uncertainties in the knowledge of the spacecraft orbits and armlengths propagate into the GW response, and are they large enough to affect LISA science?}
The inter-spacecraft arm lengths can be monitored to high precision via onboard interferometric ranging. However, the absolute positions of the spacecraft in the solar-system barycentric frame are determined from ground-based radiometric tracking, which introduces positional uncertainties of order $50$~km~\cite{2021JAnSc..68..402M}.
A related question concerns the LISA response mismodeling: 
\textit{
what is the impact of neglecting spacecraft velocities in the response when analysing Galactic binaries?
}
This has been rigorously quantified only for massive black hole binaries in Ref.~\cite{2025arXiv250910038V}.

In this paper, we address both questions by computing waveform mismatches between nominal and perturbed GW responses.
We consider two complementary orbital scenarios: (i) a static, Sun-centred toy model in which the LISA triangle is subjected to controlled perturbations in arm length, orientation, and position, and (ii) a realistic evolving scenario based on ESA reference orbits obtained from a high-fidelity orbit-determination simulation using the GODOT~\cite{godot} and MIDAS~\cite{midas} software packages.
For the response mismodeling study, we use a fast galactic-binary waveform model~\cite{2007PhRvD..76h3006C,2022MNRAS.517..697K,2018PhRvD..98f4012R,michael_katz_2025_16999246} and compare the full Doppler-boosted response~\cite{2025arXiv250910038V} against the zeroth-order approximation in which spacecraft velocities are neglected.
In both cases, the mismatch is evaluated across frequency, sky position, and polarization, providing a direct measure of the data-analysis impact.

We find that orbit determination uncertainties produce negligible mismatches across the LISA band, with the largest effects appearing above $10$~mHz where the response is most sensitive to absolute spacecraft positions.
Neglecting spacecraft velocities in the response model also leads to negligible mismatches, with the dominant impact at lower frequencies where velocity-dependent Doppler terms contribute most.
These results establish a quantitative foundation for assessing the robustness of LISA data-analysis pipelines against both orbit uncertainties and response approximations.
This paper is organised as follows. 
Section~\ref{sec:methods} describes the orbital models, the orbit-determination simulation, and the GW response formalism.
The mismatch metric used to quantify the impact of perturbations is also introduced there.
The results are presented in Sec.~\ref{sec:results}, followed by a discussion and conclusions in Sec.~\ref{sec:conclusions}.
The code used in this work is publicly available at \url{https://github.com/lorenzsp/ResponseRequirements}.
\section{Methods}
\label{sec:methods}
\subsection{Toy Model Orbit Uncertainties}
\label{subsect:toy_model}
\begin{figure}[h]
    \centering
    \includegraphics[width=0.9\columnwidth]{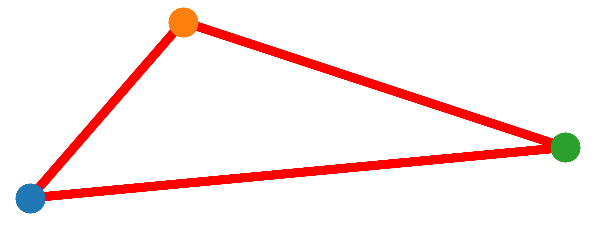}
    \caption{
    Static toy model of the LISA constellation. The three spacecraft (SC1, SC2, SC3) occupy fixed positions at the vertices of a near-equilateral triangle with nominal arm lengths of $2.5 \times 10^6$ km. Three independent classes of perturbation — arm-length variation, rigid-body rotation, and global translation — are applied sequentially to simulate spacecraft orbit uncertainties.
    }
    \label{fig:static_orbit}
\end{figure}

To isolate the geometric effects of a perturbed static constellation from the complexity of realistic orbital dynamics, we first consider a toy model in which the LISA constellation is represented as a static, triangular configuration with time-independent arm lengths (see Fig.~\ref{fig:static_orbit}).
In this simplified framework, the three spacecraft occupy constant positions in space, enabling a clean separation of the different classes of positional error.

The perturbation model is designed to reflect a key feature of the LISA orbit determination problem: the inter-spacecraft arm lengths are known to high precision from the onboard interferometric ranging system, whereas the absolute positions of the spacecraft in a solar-system barycentric frame are determined with considerably lower accuracy by ground-based radiometric tracking. 

Accordingly, instead of randomly perturbing each individual spacecraft position, we decompose the problem into three classes of perturbations:
\begin{figure*}
    \centering
    \includegraphics[]{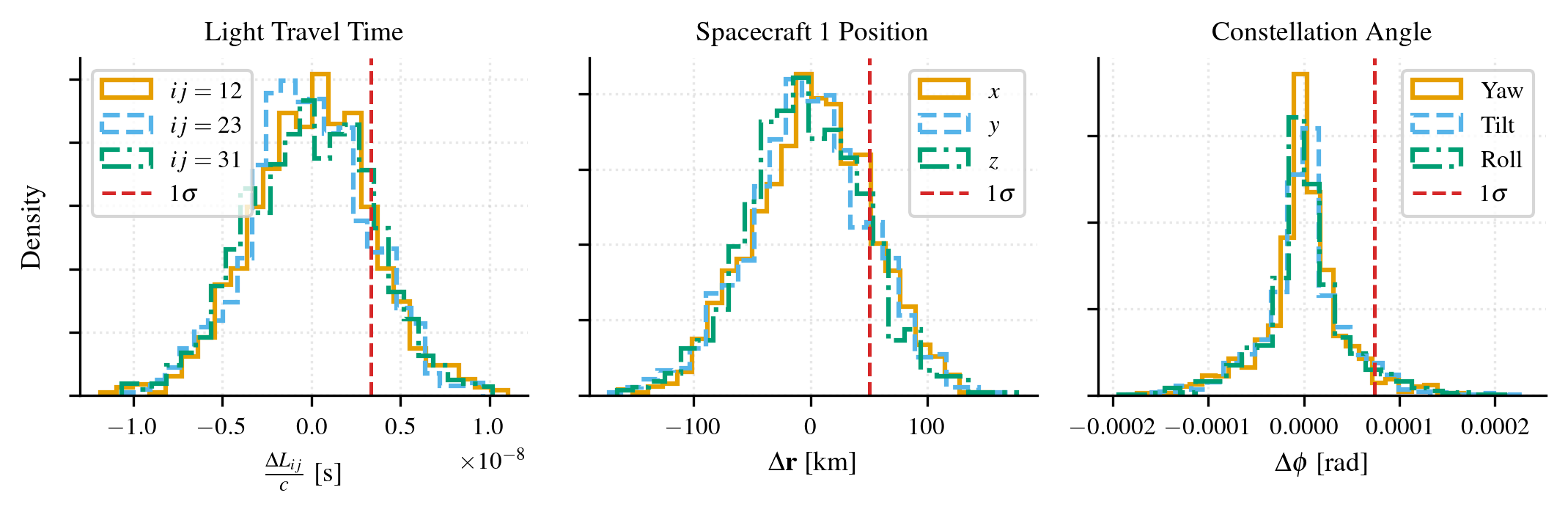}
    \caption{
    Distributions of constellation uncertainties in the static toy model (Fig. \ref{fig:static_orbit}) for the three perturbation classes. \textit{Left}: one-way light-travel time residuals $\Delta L_{ij}/c$ for all three arm pairs obtained from arm-length perturbations. \textit{Center}: Cartesian position residuals $\Delta\mathbf{r}$ of spacecraft 1 in the three coordinate directions obtained with rigid-body translations. \textit{Right}: orientation residuals in yaw, tilt, and roll angles obtained with rigid-body rotations. Dashed red vertical lines indicate the $1\sigma$ levels chosen in this work.
    }
    \label{fig:static_residuals}
\end{figure*}

\paragraph{Armlength perturbation.}
The arm lengths $L_{ij}(t)$ between spacecraft $j$ and spacecraft $i$ are usually quantified in terms of the light travel time (LTT) $L_{ij}(t)/c$ and measured with dedicated pseudoranging\footnote{As the name suggests, these pseudoranging measurements {do not} directly measure the LTT in a global reference frame, but also contain information on the mismatch of the respective spacecraft clocks. This clock information can be disentangled from the LTT in post-processing~\cite{Wang:2014zba,Wang:2015kja,Reinhardt:2024plf}. For the scope of this section, we simply assume this process is able to recover the constellation geometry to the aforementioned accuracy.} measurements~\cite{Sutton2010,Heinzel2011,Esteban2011,Yamamoto:2024rgn}.
Each of the nominal arm lengths $L_{ij} = 2.5\times10^{6}$~km is independently perturbed by an additive {zero-mean} Gaussian deviate,
\begin{equation}
    L_{ij} \to L_{ij} + \delta L_{ij}, \qquad \delta L_{ij} \sim \mathcal{N}(0,\,\sigma_L^2),
\end{equation}
where $\sigma_L$ denotes the arm length knowledge error set to $\sigma_L=1$ meter.
A new static constellation is then constructed from the perturbed arm lengths.

\paragraph{Rigid-body rotation.}\label{par:rotations}
The constellation is allowed to undergo a rigid-body rotation about a randomly oriented axis $\vu{m}$ drawn uniformly on the unit sphere, by an angle $\phi$. The rotation angle is set by requiring that it induces a positional uncertainty equivalent to a characteristic displacement error $\sigma_{\rm rot} = 50\,\mathrm{km}$. 

In the small-angle approximation, a rotation by $\phi$ induces a displacement of order $\bar{r}\,\phi$ for a spacecraft located at a distance $\bar{r}$ from the constellation centroid. Matching this to the desired $1-\sigma$ displacement leads to
\[
\phi = \frac{3}{\sqrt{2}}\,\frac{\sigma_{\rm rot}}{\bar{r}}.
\]
The numerical prefactor is chosen such that the resulting positional uncertainties are comparable to those obtained from the translational perturbations discussed below; further details are provided in Appendix~\ref{app:rotation-error-scaling}.

The rotation is implemented using Rodrigues' formula,
\begin{equation}\label{eq:rodrigues}
    \vb{T} = \vb{I} + \sin\phi\,\vb{K} + (1-\cos\phi)\,\vb{K}^2,
\end{equation}
where $\vb{T}$ is the rotation matrix corresponding to a counterclockwise rotation by angle $\phi$ about $\vu{m}$, and
\begin{equation}
    \vb{K}= 
\left[\begin{array}{rrr}
0 & -m_z & m_y \\
m_z & 0 & -m_x \\
-m_y & m_x & 0
\end{array}\right]
\end{equation}
is the skew-symmetric matrix associated with $\vu{m}$. The perturbed position of spacecraft $i$ is then given by $\vb{r}_i \rightarrow \vb{T}\,\vb{r}_i$.

\paragraph{Rigid-body translation.}\label{par:translations}
Finally, a common translational offset is applied to all three spacecraft positions,
\begin{equation}
    \vb{r}_i \to \vb{r}_i + \boldsymbol{\delta},\qquad \delta_j \sim \mathcal{N}(0,\,\sigma_{\rm trans}^2),\quad j=x,y,z,
\end{equation}
where $\sigma_{\rm trans} = 50 \,\mathrm{km}$  sets the magnitude of the absolute position uncertainty.
Since the translation is identical for all spacecraft, the arm lengths and the relative geometry of the constellation are exactly preserved.

Separating these three effects allows us to isolate the impact of each perturbation class independently. Then, we can isolate the effect of rotations and translations, which we expect to affect source uncertainties in different ways. Since LISA's antenna pattern is anisotropic, rotations of the constellation can directly translate into both phase and amplitude errors in the modeled source. A pure translation of the constellation, on the other hand, only affects the time of arrival of a given signal at the constellation, and should therefore manifest as a pure phase shift.

Independently applying these three operations produces perturbed constellations that leave response-knowledge errors shown in Fig.~\ref{fig:static_residuals} for the one-way light-travel time, the positions, and the yaw, tilt and roll angles as defined in Appendix~\ref{sec:yaw_tilt_roll}.
The inter-spacecraft distances change only at the meter level (left panel of Fig.~\ref{fig:static_residuals}), 
while the absolute position and orientation of the constellation (yaw, tilt and roll angles) are uncertain at the $\sim\!50$~km level (center and right panels of Fig.~\ref{fig:static_residuals}), 
within the expected accuracy of the LISA orbit determination from ground tracking~\cite{2021JAnSc..68..402M}.

\subsection{Realistic Orbit Uncertainties}
\begin{figure}[h]
    \centering
    \includegraphics[width=0.9\columnwidth]{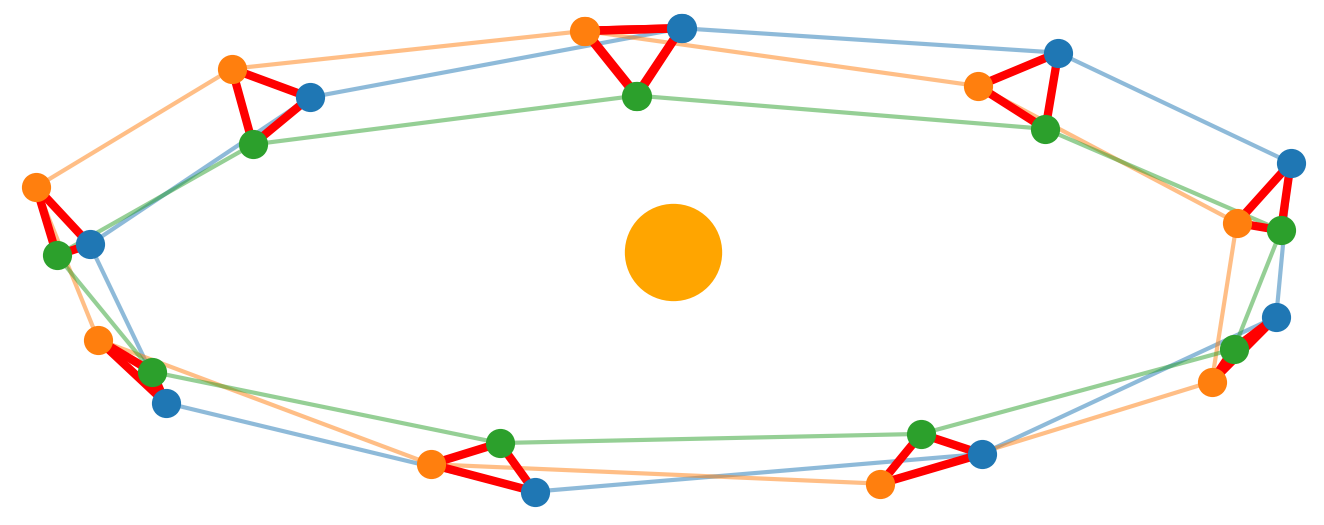}
    \caption{Realistic evolving orbit simulation based on ESA reference models. The three spacecraft (SC1, SC2, SC3) follow heliocentric trajectories that maintain a near-triangular formation with arm lengths of approximately 2.5 million kilometers. 
    The simulation incorporates perturbations from self-gravity, non-gravitational forces, and clock noise, as well as realistic measurement models for radiometric tracking and inter-spacecraft ranging.}
    \label{fig:evolving_orbit}
\end{figure}

The simulation pipeline for realistic orbits is structured in two sequential stages. 
The first stage establishes a high-fidelity truth model of the LISA constellation, generates synthetic observations, and runs the augmented orbit determination filter.
The second stage loads the filter output and generates an ensemble of trajectory estimates by drawing samples of orbit realizations from the estimated orbit covariance.

\subsubsection{Orbit Determination Simulation}

The simulation is built on ESA's GODOT~\cite{godot} astrodynamics library and the MIDAS~\cite{midas} package for additional orbit determination functionality.
The simulator separates the \textit{estimated} trajectory from the \textit{real-world} (RW) trajectory: the estimated trajectory represents the on-filter nominal model, while the real-world trajectory plays the role of the ground-truth spacecraft whose states are unknown to the filter.
A Square-Root Information Filter (SRIF) processes the observations and estimates the trajectory.

The reference LISA orbit~\cite{2021JAnSc..68..402M} is used as the nominal trajectory.
The equations of motion include point-mass solar-system gravity (JPL planetary ephemerides), together with two contributions from self-gravity-induced accelerations (SGIA): a deterministic linearly time-varying component (amplitude $\pm 2\times10^{-12}$~km\,s$^{-2}$) and a stochastic component representing unmodelled residual perturbations (uniform draws from $\pm 10^{-12}$~km\,s$^{-2}$ per axis).
Non-gravitational accelerations (NGA) are modeled as an exponentially correlated random variable (ECRV) process with $\sigma_\text{NGA} = 10^{-14}$~km\,s$^{-2}$ and correlation time $\tau_\text{NGA} = 7$~days.

Each spacecraft clock is modeled as a polynomial with three frequency-stability coefficients and an initial offset, with 1-$\sigma$ values ($5\times10^{-7}$~s, $1.6\times10^{-14}$~s\,s$^{-1}$, $9\times10^{-23}$~s\,s$^{-2}$) consistent with LISA-class oscillators~\cite{Hartwig2022}.
A stochastic clock noise with one-sided power spectral density~\cite{Hartwig2022}
\begin{equation}
    S_{\mathrm{clock}}(f) = (10^{-14})^2\left[\left(\frac{f}{1\,\mathrm{Hz}}\right)^{-1}+\left(\frac{f}{1\,\mathrm{Hz}}\right)^{-3}\right]\,\mathrm{Hz}^{-1}
\end{equation}
is added to the real-world clock, while the estimated clock is augmented with an ECRV process noise.

Four observation types are simulated over a 30-day observational arc: two-way range, two-way Doppler, inter-spacecraft pseudoranges, and time-couple measurements from a ground station.
The noise assumptions are summarized in Table~\ref{tab:observation_assumptions}.
Each observation is corrupted by additive white Gaussian noise plus a fixed random bias representing unmodelled systematic errors.
\begin{table}[h]
    \centering
    \setlength{\tabcolsep}{3.5pt}
    \begin{tabular}{c|c|c|c|c}
                         & Random noise  & Sampling time & Bias  \\
                         & $\sigma$-value  &               &       \\
        \hline
         Range           & 2m            & 1h            & 10m  \\
         Doppler         & 0.1mm/s       & 10min         & None \\
         Time Couple    & 0.05ms        & 1h            & None  \\
         pseudorange    & 1m            & 1h            & 0.1m  \\
    \end{tabular}
    
    \caption{Noise assumptions for the four observation types used in the orbit determination simulation.
    \textit{Range}: two-way ground-to-spacecraft ranging, measuring the round-trip light travel time and hence the absolute spacecraft distance from the ground station.
    \textit{Doppler}: two-way Doppler shift of the ground-uplink carrier, measuring the line-of-sight spacecraft velocity relative to the ground station.
    \textit{Time Couple}: a clock synchronisation measurement between a ground station and a spacecraft, providing an absolute epoch tie between the on-board oscillator and the ground time reference.
    \textit{Pseudorange}: inter-spacecraft one-way ranging obtained from the on-board metrology chain, measuring the one-way light travel time between two spacecraft and thus directly constraining the arm lengths.
    For each type, the table lists the 1-$\sigma$ white noise level $\sigma$, the measurement sampling interval, and the fixed systematic bias applied to the real-world observations.}
    
    \label{tab:observation_assumptions}
\end{table}

The real-world trajectory is perturbed from the nominal by displacing each spacecraft's initial state along the Radial--Along-track--Cross-track (RAC) frame, with 1-$\sigma$ values of $10/3$~km radial, $100/3$~km cross-track, and $10/3$~km along-track for position, and $10/3$, $50/3$, $10/3$ ($\times 10^{-6}$~km\,s$^{-1}$) for velocity, reflecting realistic cartwheel insertion uncertainties~\cite{2021JAnSc..68..402M}.
The SRIF processes observations sequentially and updates the estimated trajectory.
The estimated parameters comprise the spacecraft initial states, clock model coefficients and process noise, self-gravity-induced accelerations (SGIA), non-gravitational accelerations (NGA), ground station position, and measurement biases.
After convergence, the orbit covariance matrix is the primary input to the trajectory sampling stage.

\subsubsection{Trajectory Sampling}

Parameter vectors are drawn from the orbit determination posterior using the considered orbit covariance.
Each sample is propagated using the same GODOT dynamical models to produce a statistically consistent trajectory realization.
The one-way light-travel time (LTT) for all six ordered spacecraft pairs -- $(1{\to}2)$, $(2{\to}1)$, $(2{\to}3)$, $(3{\to}2)$, $(3{\to}1)$, $(1{\to}3)$ -- is then computed on a uniform epoch grid, with post-Newtonian Shapiro corrections from the Sun and the Earth applied.
\begin{figure*}
    \centering
    \includegraphics[]{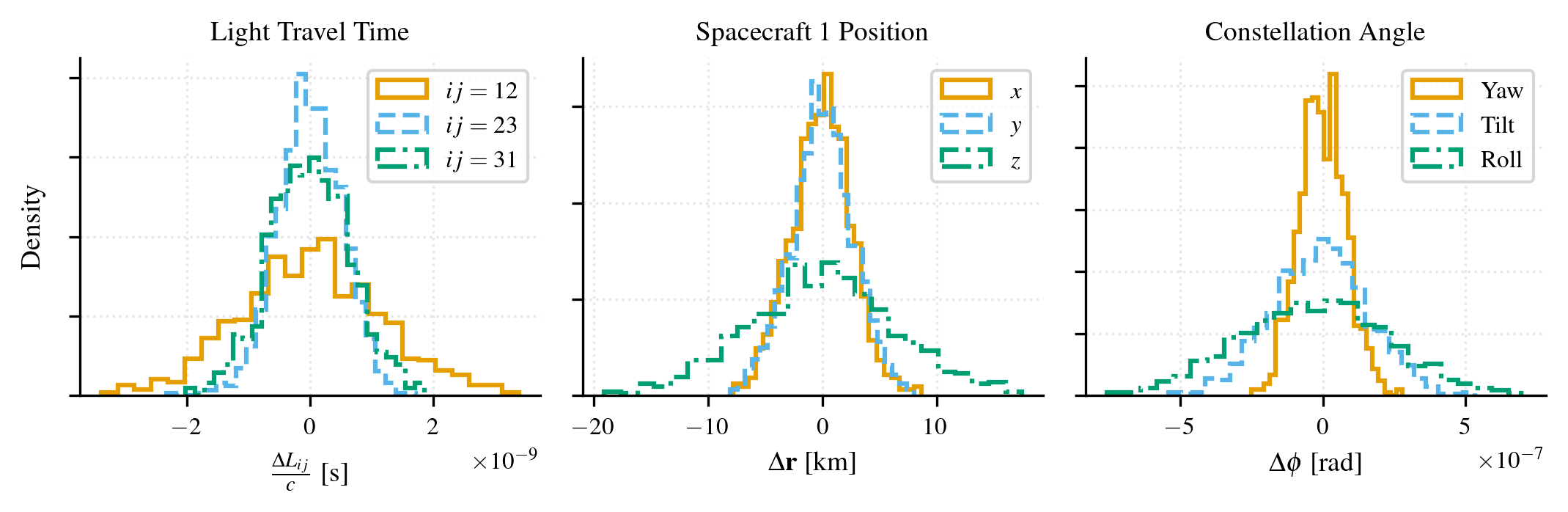}
    \caption{
        Distributions of simulated constellation uncertainties for the perturbed constellation with respect to the nominal in the realistic LISA constellation model of Fig.~\ref{fig:evolving_orbit}.
    The left panel shows the distribution of one-way light-time error, the center panel shows the distribution of position errors for the spacecraft 1 and the right panel shows the distribution of constellation plane tilt angle errors after 30  days of propagation.
    Compared to the static toy model errors in Fig.~\ref{fig:static_residuals}, arm-length errors are $\sim$3-6 times smaller, positional errors $\sim$8-20 times smaller, and angle errors $\sim$300 times smaller.
    }
    \label{fig:evolving_residuals}
\end{figure*}

The final constellation realizations produce a set of time series for the spacecraft states and the LTTs, which are used to compute the perturbed LISA response and compare it against a nominal response defined by the mean of these realizations.
The errors are shown in Fig.~\ref{fig:evolving_residuals} for the LTT and the position residuals in the Sun-centred frame and constellation tilt angle.
The standard deviations for the arm length are at most $1.13\times 10^{-9}$ seconds (about 0.34 meters), while the position errors are at most 7.12 km and the tilt angle error is at most $1.34\times 10^{-7}$ radians.
Compared to the static toy model, the arm length errors are about 3-6 times smaller, the position errors about 8-20 times smaller, and angle errors about 300 times smaller.

\subsection{LISA Response}
\label{sec:lisa_response}
This section summarises how orbital information enters the LISA detector output relevant to gravitational wave data analysis.
We follow Refs.~\cite{2023JCAP...04..066B,2025arXiv250910038V} for the derivation of the response function, and refer the reader to those works for further details and discussion.
The LISA measurement is based on one-way heterodyne interferometric links between pairs of free-falling test masses housed in the three spacecraft.
For a gravitational wave propagating along the direction $\vu{k}$, the single-link response of the link from spacecraft $j$ to spacecraft $i$ in polarization $p$ ($+$ or $\times$) is given in the time domain by~\cite{1998PhRvD..57.7089C,2003PhRvD..67b9905C,2018arXiv180610734M,2023JCAP...04..066B}
\begin{eqnarray}
\label{eq:single_link_td}
  y_{ij,p}(t, \vu{k}) &=& \frac{\xi_p(\vu{k},\vu{n}_{ij}(t))}{2(1 - \vu{n}_{ij}(t) \vdot
        \vu{k})} \nonumber \\
        && \times \left [h_p\qty(t - \frac{L_{ij}(t)}{c} - \frac{\vu{k} \vdot
        \vb{r}_j(t)}{c}) \right. \nonumber \\
        && \left. - h_p\qty(t - \frac{\vu{k} \vdot \vb{r}_i(t)}{c}) \right ]+ \mathcal{O}\qty(\qty|\frac{\vb{\dot{r}}_i}{c}|),
\end{eqnarray}
where $\vu{n}_{ij}(t) = (\vb{r}_i(t) - \vb{r}_j(t))/L_{ij}(t)$ is the unit vector pointing from spacecraft $j$ to spacecraft $i$, $\vb{r}_i(t)$ and $\vb{\dot{r}}_i(t)$ denote the position and velocity of spacecraft $i$ in the solar-system barycentric (SSB) frame, $L_{ij}(t)$ is the arm length, $h_p$ is the strain at the SSB in polarization $p$, and $c$ is the speed of light.
The two terms in brackets represent the gravitational wave phase sampled at opposite ends of the arm, evaluated at the retarded times appropriate for a photon travelling from $j$ to $i$.

The antenna pattern functions $\xi_p(\vu{k},\vu{n}_{ij}(t))$ encode the projection of the gravitational wave tensor onto the link direction and are defined, in the linear (plus and cross) polarization basis, as
\begin{align}
\label{eq:antenna_pattern}
    \xi_+ &= \qty(\vu{n}_{ij} \vdot \vu{u})^2 - \qty(\vu{n}_{ij} \vdot \vu{v})^2, \\
    \xi_\times &= 2 \qty(\vu{n}_{ij} \vdot \vu{u}) \qty(\vu{n}_{ij} \vdot \vu{v}),
\end{align}
where $\vu{u}$ and $\vu{v}$ are orthonormal vectors spanning the plane transverse to $\vu{k}$ that define the polarization frame.

Using the Fourier representation of the gravitational wave strain,
\begin{eqnarray}
  h_p(t) &=& \int_{-\infty}^{\infty} \tilde{h}_p(f') \, e^{2\pi i f' t} \, \dd f',\nonumber \\
\end{eqnarray}
and applying the Fourier transform to Eq.~\eqref{eq:single_link_td}, we obtain the time- and frequency-domain single-link response~\cite{2023JCAP...04..066B}:
\begin{eqnarray}
        \tilde{y}_{ij,p}(f, \vu{k}) &=& \iint G_{ij,p}(f',t)  \tilde{h}_p(f') e^{2 \pi i f't} \, \dd f' \, e^{-2 \pi i f t} \dd t \nonumber \\
        y_{ij,p}(t, \vu{k}) &=& \int G_{ij,p}(f',t)  \tilde{h}_p(f') e^{2 \pi i f't} \, \dd f' \label{eq:td_single_link} \\
        \tilde{y}_{ij,p}(f, \vu{k}) &=& \int \tilde{G}_{ij,p}(f',f-f')  \, \tilde{h}_p(f')  \, \dd f' \label{eq:fd_single_link}.
\end{eqnarray}
where $G_{ij,p}$ is the single-link transfer function, given by
\begin{equation}
\label{eq:response_kernel}
\begin{aligned}
  G_{ij,p}(f, t, \vu{k}) &= \frac{\xi_p(\vu{k},\vu{n}_{ij}(t))}{2(1 - \vu{n}_{ij}(t) \vdot \vu{k})} \left[\mathrm{e}^{-2\pi i f \frac{L_{ij}(t) + \vu{k} \vdot \vb{r}_j(t)}{c}} \right. \\
  &\quad \left. - \mathrm{e}^{-2\pi i f \frac{\vu{k} \vdot \vb{r}_i(t)}{c}}\right]+ \mathcal{O}\qty(\qty|\frac{\vb{\dot{r}}_i}{c}|).
\end{aligned}
\end{equation}

So far we neglected the motion of the spacecraft and omitted velocity-dependent terms $\dot{\vb{r}}_i/c$, which can be taken into account using Eq.~(A17) from Ref.~\cite{2025arXiv250910038V}
and obtain:
\begin{widetext}
\begin{equation}
\label{eq:response_kernel_db}
\begin{aligned}
  G_{ij,p}(f, t, \vu{k}) \approx \frac{\xi_p(\vu{k},\vu{n}_{ij})}{2(1 - \vu{n}_{ij} \vdot \vu{k})} 
  \Bigg\{
  \mathrm{e}^{-2\pi i f \frac{L_{ij} + \vu{k} \vdot \vb{r}_j}{c}} 
  \qty[1
  -
  \frac{\vu{k} \vdot \dot{\vb{r}}_j}{c}
  + \frac{\vu{n}_{ij} \vdot \dot{\vb{r}}_i}{c}
  ] 
   - \mathrm{e}^{-2\pi i f \frac{\vu{k} \vdot \vb{r}_i}{c}}
  \qty[1
  -\frac{\vu{k} \vdot \dot{\vb{r}}_i}{c}
  + \frac{\vu{n}_{ij} \vdot \qty(\dot{\vb{r}}_j - 2 \dot{\vb{r}}_i)}{c}
  ]
  \Bigg\} 
\end{aligned}
\end{equation}
\end{widetext}
where we neglected the last term in Eq.~(A17) of Ref.~\cite{2025arXiv250910038V} which represents the modulation of the laser's Doppler shift caused by the GW lensing the laser trajectory and higher orders in velocities $\mathcal{O}\qty(\qty|\frac{\vb{\dot{r}}_i}{c}|^2 h)$.
We implement Eq.~\eqref{eq:response_kernel_db} in \texttt{segwo} \cite{bayle2025segwo} and \texttt{jaxgb} \cite{bayle2025jaxgb} for studying the impact of mismodeling due to neglecting the velocity terms $|\dot{\vb{r}_i}|/c\sim 10^{-4}$ in the response of the LISA response.

\begin{figure}[h!]
    \centering
    \includegraphics[]{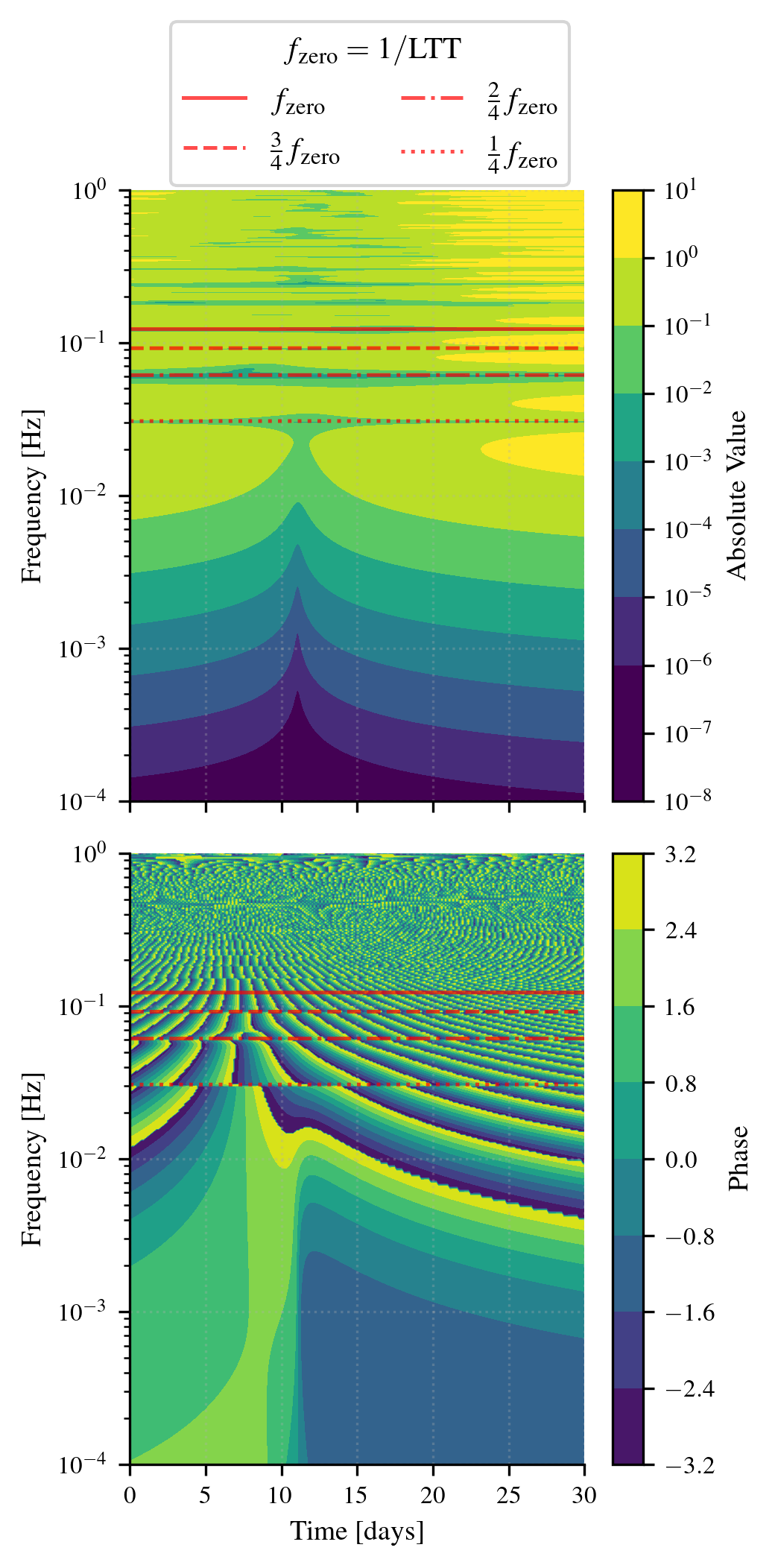}
    \caption{
    Time–frequency evolution of the LISA response function $\vb{R}(f,t)$ for the TDI A channel and polarization $h_+$, evaluated using the realistic orbital trajectories at a given sky position. We show the absolute value of the response $|\vb{R}(f,t)|$ in the \textit{top} panel and the phase of the response $\arg(\vb{R}(f,t))$ in the \textit{bottom} panel in radians.
    Some of the zeros of the response are shown as red lines and obtained by rational multiples of the inverse of the light travel time (LTT) between spacecrafts.
    The timescale over which the response varies is shown in Fig.~\ref{fig:response_time_scale}.
    }
    \label{fig:response_evolution}
\end{figure}
In what follows, we adopt the static limit in which the response kernel is constant over a given timescale discussed in Appendix~\ref{app:response_timescale} so that we can simplify
\begin{equation}
\tilde{y}_{ij,p}(f, \hat{k}) \approx G_{ij,p}(f, t, \hat{k}) \tilde h_p(f) \, .
\end{equation}
The timescale over which the response can be considered static depends on the frequency content of the gravitational wave signal and can be chosen so that the response evolution is negligible within each analysis segment.

To obtain the full interferometric response, the single-link responses are combined according to the Time-Delay Interferometry (TDI) scheme, 
which synthesizes virtual equal-arm interferometers by linearly combining time-shifted single-link measurements so as to cancel the dominant laser frequency noise~\cite{1999ApJ...527..814A,2021LRR....24....1T}.
The end-to-end computation is most conveniently expressed as a sequence of matrices that act on the gravitational wave strain.
The single-link $\tilde{y}_{ij,p}$ of Eq.~\eqref{eq:fd_single_link} is a matrix  of shape $(N_{\rm link} \times N_{\rm pol})$, for six oriented links and the $+,\times$ polarizations, which maps the two-component strain vector $(\tilde{h}_+, \tilde{h}_\times)$ to the six fractional-frequency deviations $y_{ij}$.
The TDI combination is then encoded in a second matrix $\vb{M}$ of shape $(N_{\rm TDI} \times N_{\rm link})$, 
which implements the appropriate time-delay operators for the chosen TDI generation.
In the frequency domain, each delay $\tau_{ij} \equiv L_{ij}/c$ acts as a phase shift $\mathrm{e}^{-2\pi i f \tau_{ij}}$, so $\vb{M}$ is itself frequency-dependent and is constructed directly from the one-way light-travel times.
Because light travel times evolve slowly with time, we compute chained delays as simple sums of delays instead of nested delays, which is a good approximation when computing the response function \cite{2023JCAP...04..066B}.
We adopt the orthogonal $A$, $E$, $T$ channel basis, which diagonalizes the noise covariance matrix for equal arm lengths.

For a given orbital evolution $L_{ij}, \vb{r}_i,\vb{\dot r}_i$, the
response $\vb{R}(f,t)$ with shape $N_\mathrm{TDI} \times N_{\mathrm{pol}}$ is the composition of the TDI matrix and the kernel matrix:
\begin{equation}
    \label{eq:mixing_composition}
    \vb{R}(f,t) = \vb{M}(f) \, {\vb{G}}(f,t)\, ,
\end{equation}
and the full observable, $\tilde{\vb{d}}$, is obtained by applying $\vb{R}$ to
the strain vector $\tilde{\vb{h}} = (\tilde{h}_+, \tilde{h}_\times)^\mathrm{T}$:
\begin{equation}
    \tilde{\vb{d}}(f) = \vb{R}(f,t)  \, \tilde{\vb{h}}(f)  \, .
\end{equation}

In Fig.~\ref{fig:response_evolution} we show the time–frequency evolution of the LISA response function $\vb{R}(f,t)$ for the TDI A channel and polarization $h_+$, evaluated using the realistic orbital trajectories at a given sky position.
Some of the zeros of the response are shown as red lines and obtained by rational multiples of the inverse of the light travel time between spacecrafts.
In the long-wavelength limit $2 \pi L_{ij} f \ll 1$, the absolute value of the response grows linearly with frequency as shown in the top panel of Fig.~\ref{fig:response_evolution}.
As the orbits evolve the absolute value of the response is modulated by the antenna pattern functions which vary by an order of magnitude over the 30 days time-scale considered here.
The bottom panel shows the phase evolution of the response which varies more quickly with increasing frequency.

\subsection{Metrics}
To assess the impact of uncertainties and mismodeling on the gravitational wave response, we compare a nominal response $\vb{R}_{\mathrm{nom}}$ with a response which has a deviation from the nominal due to mismodeling or orbit uncertainties $\vb{R}_\mathrm{dev}$. 
We define the relative amplitude change in the response as
\begin{equation}
\label{eq:relative_absolute_amplitude}
    \epsilon_\mathrm{amplitude} = \qty| \frac{|\vb{R}_{\mathrm{nom}}| - |\vb{R}_\mathrm{dev}|}{\langle |\vb{R}_{\mathrm{nom}}|\rangle_{\vu{k}}} |\, ,
\end{equation}
where the normalization is given by the average of the nominal response over the sky $\langle |\vb{R}_{\mathrm{nom}}|\rangle_{\vu{k}}$ to obtain a dimensionless quantity and avoid divergences when the response is close to zero.
We also define the absolute phase change in the response as
\begin{equation}
\label{eq:absolute_phase}
    \epsilon_\mathrm{phase} = \qty|\arg(\vb{R}_{\mathrm{nom}}) - \arg(\vb{R}_\mathrm{dev}) |\, .
\end{equation}
These quantities can be used to quantify the instantaneous response at any given time and frequency.

However, the ultimate impact of uncertainties in the LTTs and spacecraft positions and velocities affect the gravitational wave measurements through the response function, which is used to recover the signal from the data.
Under the assumption of stationary, ergodic, and Gaussian noise, the data are analyzed using the log-likelihood
\begin{equation}
\log \mathcal{L} \propto -\frac{1}{2} \, \braket{\vb{d} - \vb{s}}{\vb{d} - \vb{s}} \, ,
\end{equation}
where $\vb{d}$ denotes the time-domain data, and $\vb{s}$ the time-domain template signal, and the inner product is given by
\begin{equation}
    \label{eq:inner_product}
\braket{\vb{a}}{\vb{b}} = 4 \, \Re \int_{0}^{\infty} \tilde{\vb{a}}^{\dagger}(f) \, \vb{C}^{-1}(f) \, \tilde{\vb{b}}(f) \, \mathrm{d}f \, .
\end{equation}
Here, the noise covariance\footnote{We distinguish between the orbit covariance matrix, describing spacecraft uncertainties, and the noise covariance matrix, describing stochastic fluctuations of the detector noise.} matrix $\vb{C}(f)$ is a positive-definite matrix describing the detector noise properties, including correlations between different TDI channels. 
It is constructed from the mixing matrix $\vb{M}(f)$ and the single-link noise covariance matrix $\vb{N}(f)$ as
\begin{equation}
\vb{C}(f) = \vb{M}(f) \, \vb{N}(f) \, \vb{M}^\dagger(f) \, ,
\end{equation}
and therefore depends only on the light travel time quantities $L_{ij}$ through the mixing matrix $\vb{M}(f)$
(see Ref.~\cite{2023JCAP...04..066B} for further details).
We assume that noise at different frequencies is uncorrelated, implying that $\vb{C}(f)$ is diagonal in frequency space, but correlated between TDI channels.
The time-varying noise covariance matrix is much more impacted by the arm length evolution (up to 10\% in strain sensitivity variation) due to the breathing of the constellation than the impact of orbital uncertainties (up to $10^{-7} \, \%$ in strain sensitivity) going in the mixing matrix.

In our study the data are taken to be noiseless detector output, so in the frequency domain we have $\tilde{\vb{d}}(f)=\vb{R}_{\mathrm{nom}} (f)\tilde{\vb{h}}(f)$.
The template signal is given by $\tilde{\vb{s}}(f) = \vb{R}_{\mathrm{dev}}(f) \tilde{\vb{h}}(f)$. 
In both data and template, the input strain $\vb{h}(f)$ is assumed to be the same.
If the data $\vb{d}$ is approximately equal to the template $\vb{s}$ so that their signal-to-noise ratio is approximately equal $\rho^2 = \braket{\vb{d}}{\vb{d}} \approx \braket{\vb{s}}{\vb{s}}$, the log-likelihood
can be approximated as
\begin{equation}
\log \mathcal{L} \propto -\rho^2 \, (1 - \mathcal{M}) \, ,
\end{equation}
where the mismatch between two signals is defined as
\begin{equation}
\label{eq:mismatch}
\mathcal{M} = 1 - \frac{\braket{\vb{d}}{\vb{s}}}{\sqrt{\braket{\vb{d}}{\vb{d}} \, \braket{\vb{s}}{\vb{s}}}} \, ,
\end{equation}
and quantifies the fractional loss in the squared signal-to-noise ratio incurred when using an incorrect template.

Typically, the mismatch is maximized over relative time and phase shifts between the signals, which introduces an additional maximization step. 
For simplicity, we do not include this optimization here, and we study the special case of a single frequency strain,
\begin{equation}
    \label{eq:single_frequency_strain}
\tilde{\vb{h}}(f) = \vb{I} \, \delta(f - f_0) \, ,
\end{equation}
propagating in the direction $\vu{k}$ with sky coordinates $(\beta, \lambda)$
so that the inner product reduces to
\begin{equation}
\braket{\vb{d}}{\vb{s}} =
4 \, \Re \;
\mathrm{Tr} \Bigl[
\vb{R}_{\rm nom}^\dagger(f_0, \vu{k}) \,
\vb{C}^{-1}(f_0) \,
\vb{R}_{\rm dev}(f_0, \vu{k})
\Bigr] \, ,
\end{equation}
where the trace is taken over the TDI channels and can be used as a proxy to quantify the mismatch due to mismodeling and uncertainties in the response evaluated at any instance of time.

For a more realistic scenario we also consider a galactic binary template using \texttt{jaxgb} \cite{bayle2025jaxgb}.
For this analysis, we use the full inner product of Eq.~\eqref{eq:inner_product} for the mismatch, and a noise covariance matrix corresponding to the nominal response averaged over time.

\section{Results}
\label{sec:results}

First, we show how velocity mismodeling and orbit uncertainties impact the amplitude and phase of the response.
Then, we focus on the orbit uncertainties and study how different types of uncertainties give rise to different mismatch behaviors across frequencies.
We compare the mismatch induced by realistic orbit uncertainties with that induced by response mismodeling for a galactic-binary waveform model.
We study the results across the frequency range $f\in [10^{-4}, 1]$ Hz and in the sky by pixelating it using the HEALPix scheme with $N_{\rm side} = 6$, which corresponds to a pixel size of $\sim 3.7$ deg. 
We compute the mean, minimum and maximum per each sky pixel, frequency and repeat over $10^3$ realizations of the toy model and realistic orbits.
The mismatches referred to in the text correspond to the mean values unless otherwise stated.
Finally, we consider the worst-case scenario across these results and perform a parameter estimation study.

\begin{figure}
    \centering
     \includegraphics[]{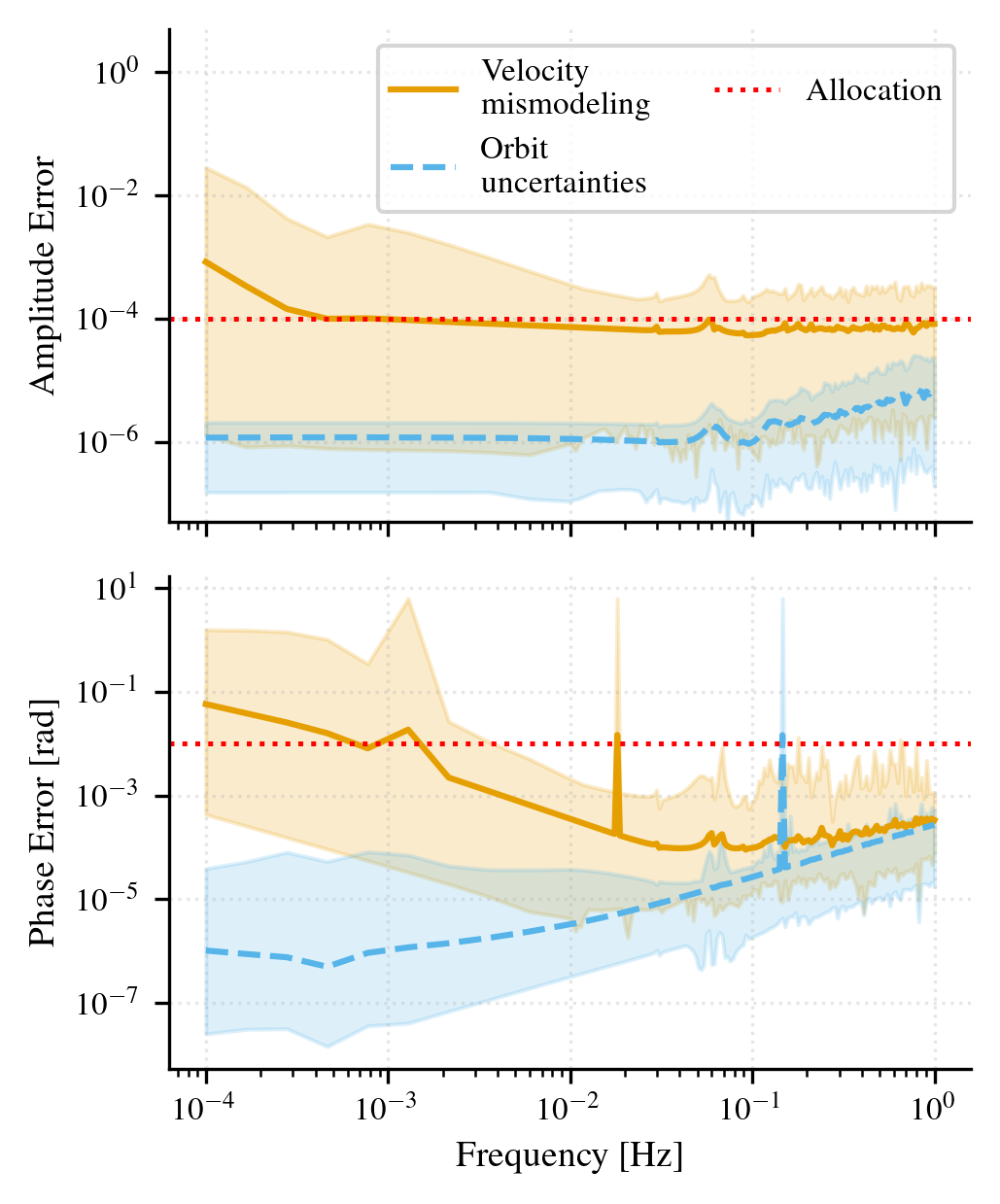}
    \caption{
     Response errors as function of frequency at 30 days for the $h_+$ polarization, TDI channel A and realistic model with errors in Fig.~\ref{fig:evolving_residuals}. 
     Solid lines show mean values and shaded bands span the minimum and maximum over sky positions and orbit realizations. 
     Orange: mismodeling error from neglecting spacecraft velocities.
     Blue dashed: orbit-determination uncertainty error.
     \textit{Top}: Error in the amplitude of the response as defined in Eq.~\ref{eq:relative_absolute_amplitude}.
     \textit{Bottom}: Error in the phase of the response as defined in Eq.~\ref{eq:absolute_phase}.
     Dotted red lines show the allocation thresholds of $10^{-4}$ (amplitude error) and $10^{-2}$ (phase error).
     }
    \label{fig:amplitude_phase_errors}
\end{figure}

\subsection{Amplitude and Phase Errors in the Response}

We quantify the impact of orbit uncertainties and velocity mismodeling through the amplitude and phase errors defined in Eqs.~(\ref{eq:relative_absolute_amplitude})--(\ref{eq:absolute_phase}). 
Figure~\ref{fig:amplitude_phase_errors} shows these quantities for the evolving response model evaluated at 30 days, where we report the relative amplitude error of Eq.~(\ref{eq:relative_absolute_amplitude}) and the absolute phase error of Eq.~(\ref{eq:absolute_phase}) for the polarization $h_+$ and for the TDI channel A.

Across the full frequency range, velocity mismodeling produces significantly larger response errors than orbit uncertainties. 
For the amplitude, the error from velocity mismodeling is $8.34\times10^{-4}$ at the lowest frequency, decreases to $9.85\times10^{-5}$ at $1\,\mathrm{mHz}$, and then remains nearly flat, with values $7.29\times10^{-5}$ at $10\,\mathrm{mHz}$ and $8.22\times10^{-5}$ at $1\,\mathrm{Hz}$. 
These errors are consistent with the velocity-to-lightspeed ratio $|\vb{\dot r}/c|\sim 10^{-4}$.
By contrast, the amplitude error due to orbit uncertainties stays at the $\sim10^{-6}$ level over most of the band, from $1.17\times10^{-6}$ at the lowest frequency to $1.18\times10^{-6}$ at $1\,\mathrm{mHz}$ and $1.12\times10^{-6}$ at $10\,\mathrm{mHz}$, before increasing to $6.43\times10^{-6}$ at $1\,\mathrm{Hz}$. 
Thus, velocity mismodeling dominates the amplitude error budget by roughly two orders of magnitude at low and intermediate frequencies, with ratios of about $84$ at $1\,\mathrm{mHz}$ and $65$ at $10\,\mathrm{mHz}$.

A similar hierarchy is observed in the phase. 
The phase error induced by velocity mismodeling is $5.79\times10^{-2}$ at the lowest frequency and $1.27\times10^{-2}$ at $1\,\mathrm{mHz}$, after which it drops sharply to $3.54\times10^{-4}$ at $10\,\mathrm{mHz}$ and remains essentially constant at $3.36\times10^{-4}$ at $1\,\mathrm{Hz}$. 
In contrast, the phase error due to orbit uncertainties is negligible at low frequency, remaining near $10^{-6}$ at the lowest frequency and at $1\,\mathrm{mHz}$, and only becoming appreciable at high frequency, where it reaches $3.29\times10^{-6}$ at $10\,\mathrm{mHz}$ and $2.73\times10^{-4}$ at $1\,\mathrm{Hz}$. 
At low frequency, velocity mismodeling therefore dominates the phase error by more than four orders of magnitude, whereas at the highest frequency shown the two contributions become comparable.

The errors due to velocity mismodeling increase towards low frequencies, approximately following an $f^{-1}$ behavior before flattening at higher frequencies. 
The errors induced by orbit uncertainties remain nearly constant at low frequency and grow only towards the high-frequency end of the band. 
These numerical trends are consistent with the scalings obtained in Appendix~\ref{sec:scaling}. 

The LISA response error budget allocation is shown in Figure~\ref{fig:amplitude_phase_errors} with dotted red lines at $10^{-4}$ for the amplitude error and at $10^{-2}$ for the phase error.
Within this comparison, velocity mismodeling exceeds or approaches these reference values at low frequencies, while orbit uncertainties remain comfortably below them over most of the band.

\subsection{Dissecting the Impact of Orbit Uncertainties}

\begin{figure*}
    \centering
    \includegraphics[]{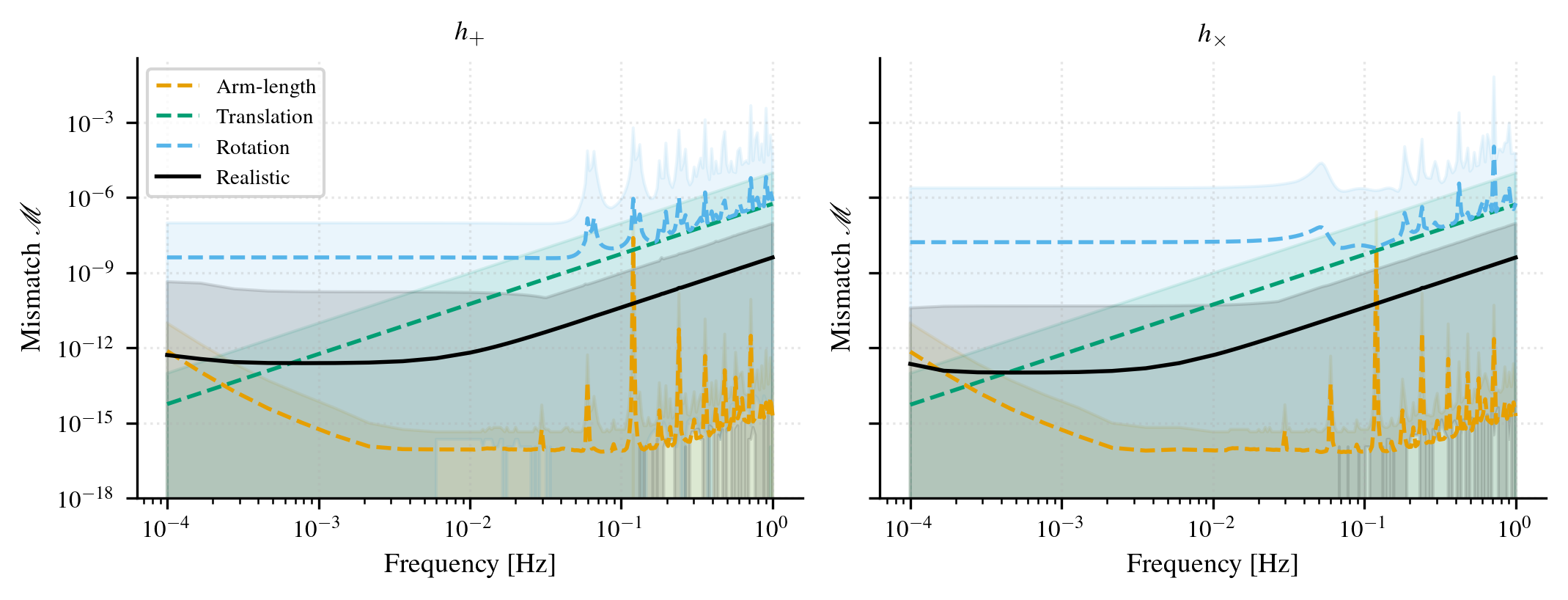}
    \caption{
     Mismatch $\mathcal{M}$ between the nominal and perturbed response as a function of frequency for the static toy model and the realistic evolving scenario.
     The mismatch is computed for a monochromatic source with mean over sky coordinates for the two polarizations and min and max over sky coordinates and orbit realizations shown as shaded regions.
     The dashed lines show the results for the static toy model, whereas the solid lines show the results for the realistic evolving scenario.
     The mismatch also exhibits oscillations as a function of frequency, which are related to the zeros of the sinc-based transfer function in the single-link response.
     }
    \label{fig:mismatch_response_model}
\end{figure*}

The mismatch $\mathcal{M}$ between the nominal and perturbed responses for both the static toy model and the realistic evolving scenario using the model of Eq.~\ref{eq:single_frequency_strain} is shown in Fig.~\ref{fig:mismatch_response_model}.

In the static toy model, the mismatch is dominated by rigid-body rotations. 
Across the frequency band, rotational perturbations produce a nearly constant mismatch of $\mathcal{M} \sim 10^{-9}$--$10^{-8}$ (e.g., $4.1\times10^{-9}$ at $1\,\mathrm{mHz}$ for $h_+$ and $1.66\times10^{-8}$ for $h_\times$), with only a mild increase at high frequencies. 
Translational perturbations are subdominant but strongly frequency-dependent, increasing from $\sim 10^{-13}$ at $1\,\mathrm{mHz}$ to $\sim 10^{-9}$ at $100\,\mathrm{mHz}$, consistent with a $\propto f^2$ scaling. 
Arm-length perturbations are negligible over most of the band, remaining at $\sim 10^{-16}$ at $1\,\mathrm{mHz}$ and not exceeding $\sim 10^{-13}$ even at the lowest frequencies.
The overall values of the mismatch depend directly on the orbit uncertainties set for the rigid-body rotations and translations $\sigma_{\rm trans}=\sigma_{\rm rot} = 50 \, \mathrm{km}$ and arm-length $\sigma_L=1 \, \mathrm{m}$ perturbations.

The realistic evolving scenario (evaluated after 30 days) exhibits lower mismatches than the static one across the full band.
This reduction reflects the smaller realistic orbit uncertainties (Fig.~\ref{fig:evolving_residuals}) than in the static toy case (Fig.~\ref{fig:static_residuals}).
At low frequencies, the mismatch is $\mathcal{M} \sim 10^{-13}$ (e.g., $2.46\times10^{-13}$ for $h_+$ at $1\,\mathrm{mHz}$), representing a reduction of approximately four orders of magnitude relative to the rotation-dominated static case. 
At higher frequencies, the mismatch increases to $\sim 10^{-11}$ at $100\,\mathrm{mHz}$, which is still two to three orders of magnitude smaller than the translation-driven mismatch of the static model at the same frequency. 

While the overall mismatches depend on the specific values of the constellation uncertainties, $\sigma_{\rm trans},\sigma_{\rm rot}, \sigma_L$, the frequency dependency of the mismatch is determined by the type of perturbation.
The static toy model is therefore a useful diagnostic tool for isolating the scaling behavior of individual perturbations. 
Arm-length perturbations are strongly suppressed and rise at low frequencies, with mismatch values $\lesssim 10^{-16}$ at $\sim1\,\mathrm{mHz}$.
Rotational perturbations provide an approximately frequency-independent contribution below the first transfer-function zero ($\sim 1/16$ Hz), and remain the dominant source of mismatch over most of the band;
Translational perturbations scale approximately as $\propto f^{2}$, increasing by several orders of magnitude from low to high frequency and becoming comparable to rotations at the upper end of the band.

The realistic evolving scenario combines these effects with significantly reduced amplitudes, reflecting the improved fidelity of the orbit reconstruction. 
As a result, the mismatch remains well below $10^{-7}$ across most of the band, indicating that orbit-related errors are strongly suppressed in a realistic implementation compared to the idealised static perturbations.
The oscillatory features visible in Fig.~\ref{fig:mismatch_response_model} arise from the zeros of the response across frequency.

\subsection{Galactic Binary Mismodeling vs Orbit Uncertainties}

\begin{figure}
    \centering
    \includegraphics[width=0.9\columnwidth]{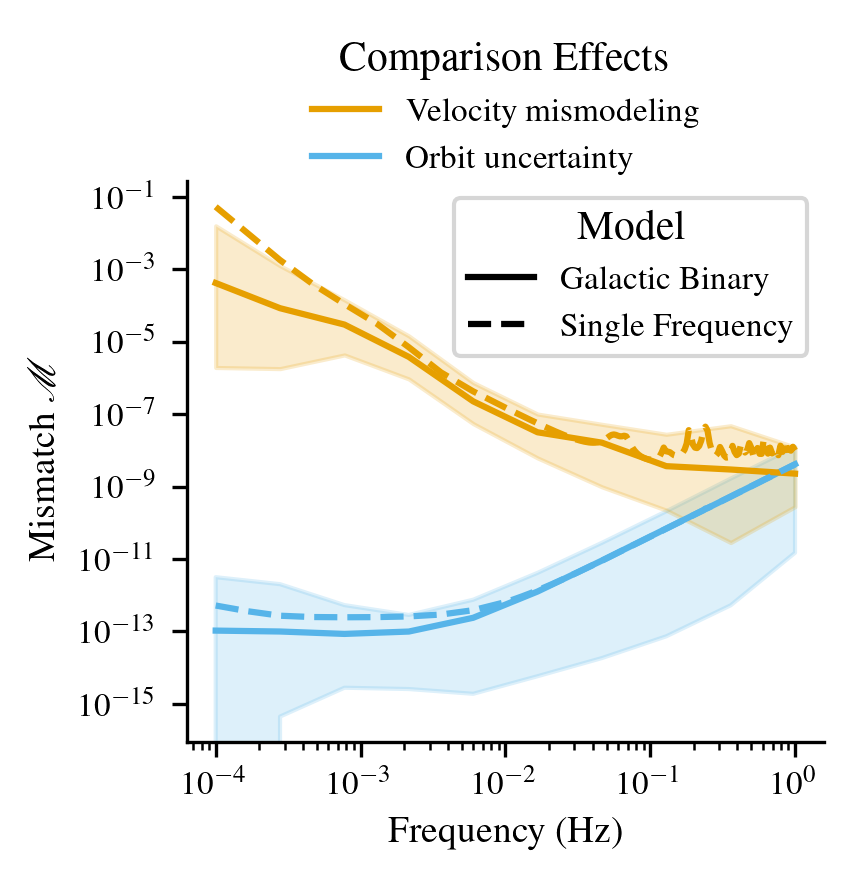}
    \caption{
    Mismatch $\mathcal{M}$ induced by orbit uncertainties (blue) and by velocity mismodeling (orange), for a 30-day observational window using realistic LISA orbits.
    The mismatch is computed both with the single-frequency model (dashed lines, Eq.~\eqref{eq:single_frequency_strain}) and with the galactic binary model (solid lines), representative of fast waveform models used in LISA data analysis.
    }
    \label{fig:mismatch_response_vs_gb}
\end{figure}

Finally, we compare the mismatch induced by orbit uncertainties with that arising from velocity mismodeling, using a galactic binary signal evaluated over a 30-day segment within the observational arc of the realistic simulation.

The mismatch due to velocity mismodeling is obtained by comparing the full response, including velocity-dependent terms (cf. Eq.~\eqref{eq:response_kernel_db}), with a response model in which these terms are neglected (Eq.~\eqref{eq:response_kernel}). 
The mismatch due to orbit uncertainties is computed by comparing responses generated with nominal and perturbed orbital parameters within the realistic evolving scenario.

The results are shown in Fig.~\ref{fig:mismatch_response_vs_gb}, where we report both contributions using two waveform models: the response-only model (dashed lines) implemented in \texttt{segwo}, and the galactic binary model (solid lines) implemented in \texttt{jaxgb}. 
The two implementations yield broadly consistent median trends, with the response-only model typically overestimating the mismatch by a factor of a few across most of the frequency range, except below $\sim 2\times10^{-4}\,\mathrm{Hz}$.
We stress that an exact agreement between the two implementations is not expected, since one is evaluated at an instantaneous time and frequency, while the other incorporates the time evolution of the system.

A clear hierarchy emerges between the two error sources. 
Velocity mismodeling dominates the mismatch across the entire frequency band. 
At low frequencies, the mismatch due to velocity effects is $\mathcal{M} \sim 4.2\times10^{-4}$ at $0.1\,\mathrm{mHz}$, decreasing to $2.6\times10^{-5}$ at $1\,\mathrm{mHz}$, and further to $1.5\times10^{-7}$ at $10\,\mathrm{mHz}$ and $8.1\times10^{-9}$ at $100\,\mathrm{mHz}$. 
In contrast, the mismatch induced by orbit uncertainties remains several orders of magnitude smaller, at the level of $\sim 10^{-13}$ at low frequencies (e.g., $1.05\times10^{-13}$ at $0.1\,\mathrm{mHz}$ and $8.73\times10^{-14}$ at $1\,\mathrm{mHz}$), increasing only at higher frequencies to $6.2\times10^{-13}$ at $10\,\mathrm{mHz}$ and $4.8\times10^{-11}$ at $100\,\mathrm{mHz}$.

As a result, velocity mismodeling exceeds the mismatch from orbit uncertainties by approximately nine orders of magnitude at $0.1\,\mathrm{mHz}$, and by about two orders of magnitude at $100\,\mathrm{mHz}$. 
This behavior is consistent with the trends observed in the amplitude and phase error analysis, where velocity effects dominate at low frequencies while orbit-induced effects grow towards higher frequencies.

Overall, these results indicate that neglecting velocity-dependent terms in the response constitutes the leading source of modelling error for galactic binary signals. 
In comparison, orbit uncertainties in realistic reconstructions produce a negligible contribution to the mismatch across most of the LISA band, with largest contribution at the highest frequency.

\subsection{Parameter Estimation with Velocity Mismodeling}

To assess the impact of this mismodeling on parameter inference, we perform a Markov Chain Monte Carlo (MCMC) analysis of a galactic binary signal. 
We select the set of source parameters yielding the largest mismatch across frequency and sky location in one year of simulated observations, as this timescale allows for accurate sky localisation.

We inject a signal with frequency $10^{-4}\,\mathrm{Hz}$ and signal-to-noise ratio $\mathrm{SNR}=200$ into one year of data. For simplicity, we assume a fixed noise covariance matrix over this period, as our primary goal is to quantify parameter biases arising from mismodeling; future studies should incorporate a time-dependent noise model.

\begin{figure}[h!]
    \centering
    \includegraphics[width=0.99\columnwidth]{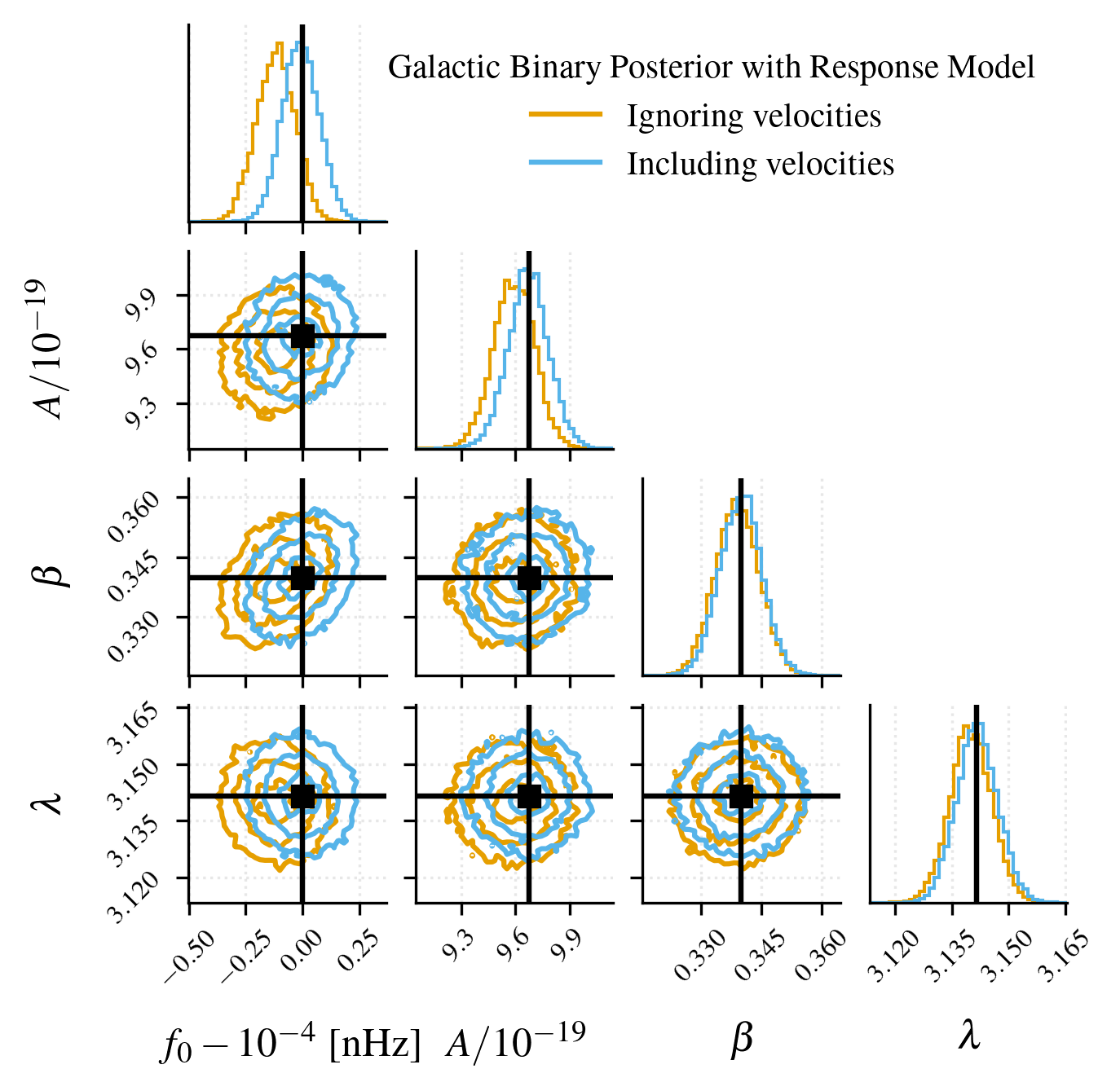}
    \caption{
    Posterior distributions for frequency $f_0$, amplitude $A$ and sky localization $\beta,\lambda$ of a galactic binary injection at $\mathrm{SNR} = 200$, recovered with two templates: one using the full response (blue, consistent with the injection) and one neglecting velocity-dependent terms (orange).
    }
    \label{fig:mcmc_gb}
\end{figure}

The resulting posteriors, shown in Fig.~\ref{fig:mcmc_gb}, are obtained using two templates: one based on the full response (blue), consistent with the injection, and one neglecting velocity-dependent terms (orange). 
We find that neglecting velocity contributions induces parameter biases at the level of less than $1\sigma$. 
In contrast, the template including the full Doppler-boosted response recovers unbiased posteriors, as expected.
The difference in mean log-likelihood between the two models is $\Delta \log \mathcal{L} \sim 10$.

\section{Discussion and Conclusions}
\label{sec:conclusions}

In this work we quantified two distinct sources of gravitational wave response error for LISA: uncertainties in spacecraft orbits and armlengths and mismodeling from neglecting velocity-dependent terms in the response kernel. 

We find that uncertainties in the knowledge of the spacecraft orbits and armlengths result in LISA detector output amplitude errors of order $10^{-5}$ (relative) and $10^{-3}$ in phase (at high frequencies).
These errors lead to mismatches at worst $10^{-7}$ unlikely to affect the LISA scientific exploitation.
By contrast, neglecting velocity-dependent terms influences the detector output especially at lower frequencies with relative amplitude of $10^{-2}$ and phase errors of one radian.
These errors lead to median mismatches of order $10^{-3}$ around $0.1$~mHz. 
The MCMC study confirms that these mismodeling errors can propagate into measurable parameter biases, including a frequency shift of about $0.6\sigma$ in the tested SNR=200 case.

These trends are supported by the analytic scaling derived in Appendix~\ref{sec:scaling}, where orbit-induced effects grow with frequency while velocity-mismodeling effects decrease with frequency. 
These latter findings are broadly consistent with previous analyses for massive black-hole binaries~\cite{2025arXiv250910038V}, and indicate that low-frequency galactic-binary studies should adopt the full Doppler-boosted response when high-fidelity inference is required since it comes at no additional computational cost.
At high frequencies the validity of the quasi-static response treatment becomes questionable, and further investigation is needed to assess the accuracy of the response modeling schemes in this regime.
In addition, the frequency-dependent zeros of the response may provide a useful calibration diagnostic for both response modeling and orbit knowledge.
Future works should extend this analysis to extreme mass ratio inspirals, stellar origin black hole binaries, stochastic backgrounds, and joint noise-response inference.


\acknowledgements
L.~S.\ is supported by the European Space Agency Research Fellowship Programme. O.H. gratefully acknowledges the support by the Deutsches Zentrum für Luft- und Raumfahrt (DLR) with funding of the Federal Ministry for Economic Affairs and Energy based on a resolution of the Deutsche Bundestag (FKZ 50OQ2301).
The authors acknowledge the use of large language models for proofreading and polishing the manuscript. 
All text was carefully reviewed and edited to ensure accuracy.

\section*{Data availability}
This work comes with publicly available \href{https://github.com/lorenzsp/ResponseRequirements}{software} to reproduce this work and the figures.

\appendix

\section{Rotation error scaling}
\label{app:rotation-error-scaling}

We derive the scaling used in Section~\ref{subsect:toy_model} to relate a small rigid-body rotation to an equivalent positional displacement of the spacecraft.

We consider a spacecraft at position $\vec r$ undergoing a rotation by a small angle $\phi$ about a random axis $\vu{m}$, uniformly distributed on the unit sphere. By rotational invariance, we may, without loss of generality, choose
\begin{equation}
    \vec r = (\bar r, 0, 0)\,.
\end{equation}
The rotation axis is parametrised as
\begin{equation}
    \vu{m} = (\sin\theta \cos\psi, \sin\theta \sin\psi, \cos\theta)\,.
\end{equation}

Using Rodrigues' formula (Eq.~\ref{eq:rodrigues}) and expanding to first order in $\phi$, the rotated vector becomes
\begin{equation}
    \vec r_p = \mathbf{T}\,\vec r = \bar r \bigl(1, \phi \cos\theta, -\phi \sin\theta \sin\psi \bigr)\,.
\end{equation}
The displacement induced by the rotation is $\Delta\vec r = \vec r_p - \vec r$, whose squared norm is
\begin{equation}
    ||\Delta \vec r||^2 = \bar r^2 \phi^2 \left( \cos^2\theta + \sin^2\theta \sin^2\psi \right)\,.
\end{equation}

Averaging over all orientations of the rotation axis yields
\begin{equation}
    \langle ||\Delta \vec r||^2 \rangle 
    = \int \frac{\sin \theta \dd \theta \dd \psi}{4\pi} \, ||\Delta \vec r||^2
    = \frac{2}{3}\,\bar r^2 \phi^2\,.
\end{equation}

For the translational perturbations discussed in Section~\ref{par:translations}, we parametrise the displacement in terms of a per-coordinate variance $\sigma_{\rm trans}^2$, such that
\begin{equation}
    \langle ||\Delta \vec r||^2 \rangle = 3\,\sigma_{\rm trans}^2\,.
\end{equation}

Matching the rotationally induced mean-squared displacement to this translational variance and identifying $\sigma_\mathrm{rot} \equiv \sigma_{\rm trans}$ yields
\begin{equation}
    \phi = \sqrt{\frac{3}{2}}\,\frac{\sqrt{\langle ||\Delta \vec r||^2 \rangle}}{\bar r}
    = \frac{3}{\sqrt{2}}\,\frac{\sigma_\mathrm{rot}}{\bar r}\,.
\end{equation}

\section{Yaw, tilt, and roll angles}
\label{sec:yaw_tilt_roll}

To characterise the orientation of the constellation, we construct a time-dependent orthonormal frame directly from the spacecraft positions. 

We define a constellation-fixed basis $(\hat{\mathbf{e}}_x, \hat{\mathbf{e}}_y, \hat{\mathbf{e}}_z)$ as follows:
\begin{itemize}
    \item $\hat{\mathbf{e}}_x$ is the unit vector pointing from spacecraft~1 to spacecraft~2,
    \item $\hat{\mathbf{e}}_z$ is the unit normal to the constellation plane, obtained from $(\mathbf{r}_2 - \mathbf{r}_1) \times (\mathbf{r}_3 - \mathbf{r}_1)$,
    \item $\hat{\mathbf{e}}_y = \hat{\mathbf{e}}_z \times \hat{\mathbf{e}}_x$ completes a right-handed orthonormal triad.
\end{itemize}

The resulting rotation matrix $\mathbf{O}$ is constructed by taking these unit vectors as its columns,
\begin{equation}
    \mathbf{O} = \bigl[\,\hat{\mathbf{e}}_x,\; \hat{\mathbf{e}}_y,\; \hat{\mathbf{e}}_z\,\bigr],
\end{equation}
which maps the body-fixed frame to the inertial frame.

We then extract the orientation of the constellation using ZYX (yaw–tilt–roll) Euler angles:
\begin{align}
    \text{yaw}  &= \arctan2(O_{10}, O_{00}), \\
    \text{tilt} &= \arcsin(-O_{20}), \\
    \text{roll} &= \arctan2(O_{21}, O_{22}).
\end{align}

These angles admit a simple geometric interpretation:
\begin{itemize}
    \item {Yaw} measures the azimuthal orientation of the arm connecting spacecraft~1 to~2 in the reference $(x,y)$ plane.
    \item {Tilt} (pitch) quantifies the inclination of the constellation plane relative to the reference plane.
    \item {Roll} describes the rotation of the triangle around its normal vector, fixing the orientation of spacecraft~3 within the plane.
\end{itemize}

\section{Response Timescale}
\label{app:response_timescale}
\begin{figure*}
    \centering
     \includegraphics[]{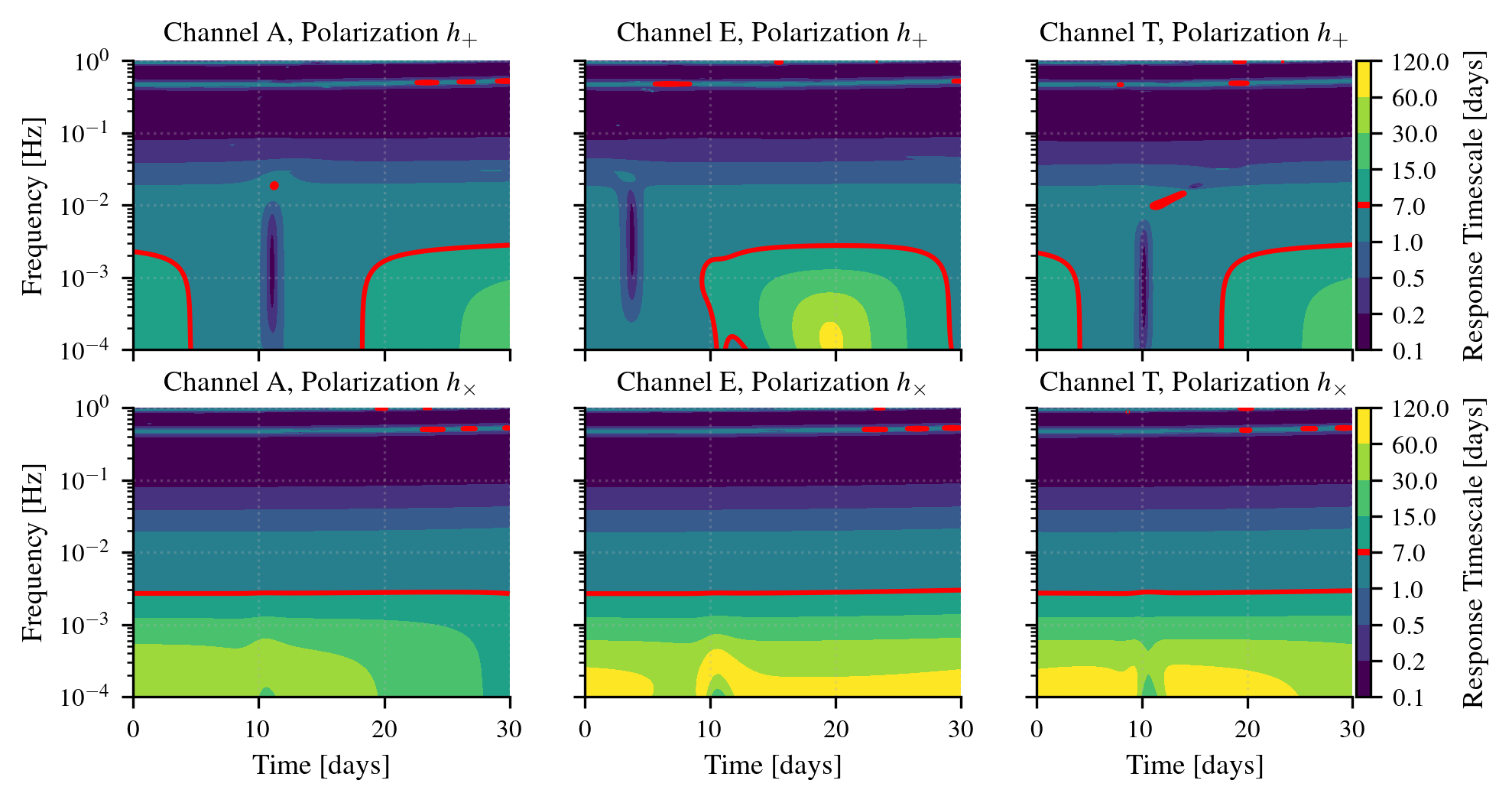}
    \caption{
    Timescale over which the LISA response varies (Eq.~\eqref{eq:timescale_response}) for 30 days of realistic orbits for the TDI channels $A,E,T$ (columns) and polarization (rows).
    The LISA response can be considered static over observational timescales smaller than the response timescale.
    For the TDI A channel and $+$ polarization the response timescale decreases rapidly around 10 days but this is due to the fact that the response goes to zero around the same time as shown in Fig.~\ref{fig:response_evolution}
     }
    \label{fig:response_time_scale}
\end{figure*}

We define the response timescale as 
\begin{equation}
\label{eq:timescale_response}
    T_r = \qty|\frac{R_{c,p}(f,t)}{\partial_t R_{c,p}(f,t)}|
\end{equation}
for each channel $c = A, E, T$ and polarisation $p = +, \times$.
The results for realistic orbits over 30 days are shown in Fig.~\ref{fig:response_time_scale}.
In general, the LISA response can be considered static when the observational timescale is shorter than $T_r$.
At frequencies below $10^{-3},\mathrm{Hz}$, the response timescale typically exceeds one week for all channels and polarisations, with local variations associated with zeros of the response, particularly for the $+$ polarisation. For example, in the TDI $A$ channel with $+$ polarisation, the response timescale drops sharply near $\sim 10$ days because the response vanishes at approximately the same time (as shown in Fig.~\ref{fig:response_evolution}).
At higher frequencies ($\gtrsim 0.1,\mathrm{Hz}$), the response varies on much shorter timescales, of order a few hours. In this regime, the observational timescale must therefore be correspondingly shorter.
A similar analysis, focused on massive and stellar-origin black hole binaries, was presented in Ref.~\cite{2018arXiv180610734M}. Here, we instead provide a source-agnostic characterisation of the response variability.

\section{Frequency Scaling of Mismatch Effects}
\label{sec:scaling}
\begin{figure}
    \centering
     \includegraphics[]{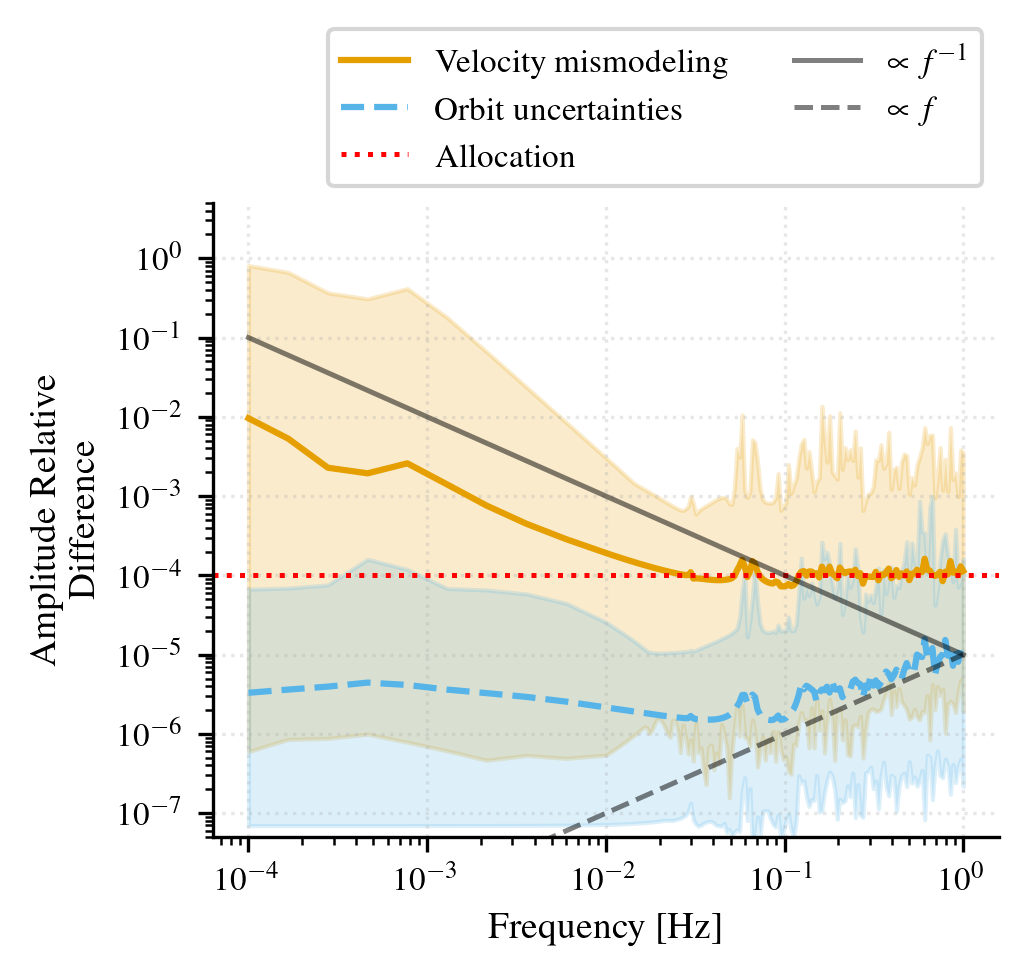}
    \caption{
     Relative error in the amplitude of the response as function of frequency at 30 days for the $h_+$ polarization, TDI channel A and realistic model with errors in Fig.~\ref{fig:evolving_residuals}. 
     Solid lines show sky-averaged mean values; shaded bands span the minimum and maximum over sky positions. 
     Orange: mismodeling error from neglecting spacecraft velocities.
     Blue dashed: orbit-determination uncertainty error.
     The shown errors differ from the one of Fig.~\ref{fig:amplitude_phase_errors} because we use Eq.~\ref{eq:rel_err_amp}.
     }
    \label{fig:amplitude_phase_errors_app}
\end{figure}
The two effects studied in this paper, orbit uncertainties and velocity mismodeling, produce errors and mismatches with opposite frequency dependences.
We derive approximate analytic scalings to explain this behavior.

We introduce compact notation for Eq.~\eqref{eq:response_kernel_db}.
Define the angular prefactor $P \equiv \xi_p / [2(1-\vu{n}_{ij}\cdot\vu{k})]$, the phase arguments
\begin{equation}
\Phi_1 \equiv \frac{2\pi f(L_{ij}+\vu{k}\cdot\vb{r}_j)}{c}, \qquad
\Phi_2 \equiv \frac{2\pi f\,\vu{k}\cdot\vb{r}_i}{c},
\end{equation}
and the dimensionless velocity corrections
\begin{align}
\varepsilon_1 &\equiv -\frac{\vu{k}\cdot\dot{\vb{r}}_j}{c} + \frac{\vu{n}_{ij}\cdot\dot{\vb{r}}_i}{c} = \mathcal{O}(v/c)\,,\\
\varepsilon_2 &\equiv -\frac{\vu{k}\cdot\dot{\vb{r}}_i}{c} + \frac{\vu{n}_{ij}\cdot(\dot{\vb{r}}_j-2\dot{\vb{r}}_i)}{c} = \mathcal{O}(v/c)\,.
\end{align}
The full Doppler-boosted kernel and its zeroth-order approximation then read
\begin{align}
G &= P\bigl[e^{-i\Phi_1}(1+\varepsilon_1) - e^{-i\Phi_2}(1+\varepsilon_2)\bigr],\\
G^{(0)} &= P\bigl[e^{-i\Phi_1} - e^{-i\Phi_2}\bigr].
\end{align}
Note that neither $P$, $\varepsilon_1$, nor $\varepsilon_2$ depends on $f$; all frequency dependence is carried by $\Phi_1$ and $\Phi_2$.

We can evaluate $|G^{(0)}|$ exactly.
Using $\vb{r}_j = \vb{r}_i - L_{ij}\vu{n}_{ij}$ one finds $\Phi_1-\Phi_2 = 2\pi f L_{ij}(1-\vu{n}_{ij}\cdot\vu{k})/c \equiv \Delta\Phi$, so
\begin{align}
|G^{(0)}| &= |P|\,\bigl|e^{-i\Phi_1}-e^{-i\Phi_2}\bigr| \\
           &= 2|P|\,\bigl|\sin\tfrac{\Delta\Phi}{2}\bigr|\\
&\xrightarrow{\;fL/c\ll 1\;} |P|\,\Delta\Phi \\ 
&= \frac{|\xi_p|\,\pi f L_{ij}}{c}\,,
\label{eq:G0_lw}
\end{align}
where in the last step the factor $(1-\vu{n}_{ij}\cdot\vu{k})$ from $\Delta\Phi$ cancels against the same factor in $|P|^{-1}$.
Thus $|G^{(0)}| \propto fL/c$ in the long-wavelength limit.

\subsection{Orbit uncertainty: $\mathcal{M}_{\rm orb}\propto f^2$}
A position translation $\delta\vb{r}$ (e.g.\ a rigid translation of the whole constellation) shifts both $\Phi_1$ and $\Phi_2$ by the same increment $\delta\psi = 2\pi f\,\vu{k}\cdot\delta\vb{r}/c$, so the perturbed response acquires a common phase factor:
\begin{equation}
G' = e^{-i\delta\psi}\,G^{(0)}, \qquad
\delta\psi = \frac{2\pi f\,\vu{k}\cdot\delta\vb{r}}{c} \propto f\,.
\end{equation}
This scaling is observed in Fig.~\ref{fig:amplitude_phase_errors_app} for the quantity
\begin{equation}
\label{eq:rel_err_amp}
    \epsilon_\mathrm{relative} = \qty| \frac{|\vb{R}_{\mathrm{nom}}| - |\vb{R}_\mathrm{dev}|}{|\vb{R}_{\mathrm{nom}}|} | \, .
\end{equation}
Since the mismatch in Eq.~\eqref{eq:mismatch} is not maximised over phase and time shifts (see Sec.~\ref{sec:methods}), this global phase offset contributes directly:
\begin{equation}
\mathcal{M}_{\rm orb}
\approx 1-\cos^2(\delta\psi)
\approx \delta\psi^2
= \left(\frac{2\pi f\,\delta r}{c}\right)^2
\propto f^2.
\label{eq:Morb_scaling}
\end{equation}
This is observed in Fig.~\ref{fig:mismatch_response_model} for the translations.

\subsection{Velocity mismodeling: $\mathcal{M}_{\rm vel}\propto f^{-2}$}
The modelling error is
\begin{equation}
\Delta G \equiv G - G^{(0)} = P\bigl[e^{-i\Phi_1}\varepsilon_1 - e^{-i\Phi_2}\varepsilon_2\bigr].
\end{equation}
Because every exponential is a unit-modulus complex number ($|e^{-i\Phi}|=1$ for all real $\Phi$, regardless of $f$), the triangle inequality gives
\begin{equation}
|\Delta G| \leq |P|\,(|\varepsilon_1|+|\varepsilon_2|) \lesssim |P|\,\frac{v}{c}\,.
\label{eq:DeltaG_bound}
\end{equation}
This bound is \textit{independent of $f$}: the $f$-dependent phases rotate $\varepsilon_1$ and $\varepsilon_2$ in the complex plane but cannot change their moduli.
The $f$-dependence of $|G^{(0)}|$, on the other hand, comes entirely from the near-cancellation of two unit-modulus exponentials, as shown in Eq.~\eqref{eq:G0_lw}.
Crucially, the prefactor $|P|$ appears in \textit{both} Eq.~\eqref{eq:DeltaG_bound} and Eq.~\eqref{eq:G0_lw} and therefore cancels in the ratio:
\begin{equation}
\frac{|\Delta G|}{|G^{(0)}|}
\lesssim \frac{|P|\,v/c}{|P|\,\Delta\Phi}
= \frac{v/c}{\Delta\Phi}
= \frac{v/c}{2\pi f L_{ij}(1-\vu{n}_{ij}\cdot\vu{k})/c}
\propto f^{-1} \, ,
\label{eq:fractional_error}
\end{equation}
which is shown in Fig.~\ref{fig:amplitude_phase_errors_app}.
The corresponding mismatch scales as
\begin{equation}
\mathcal{M}_{\rm vel}
\approx \left(\frac{v}{\,2\pi f L_{ij}(1-\vu{n}_{ij}\cdot\vu{k})\,}\right)^2
\propto f^{-2}.
\label{eq:Mvel_scaling}
\end{equation}

\bibliography{references}

\end{document}